\documentclass[sigconf, nonacm, screen]{acmart}

\usepackage{includes/packages}
\usepackage{includes/commands}
\usepackage{includes/shorthands}

\begin{document}

\title{PCCL: Process Group-Aware Scalable and Generic Collective Algorithm Synthesizer}

\author{William Won}
\email{william.won@gatech.edu}
\affiliation{
  \institution{Georgia Institute of Technology}
  \city{Atlanta}
  \state{GA}
  \country{USA}
}

\author{Kartik Lakhotia}
\email{kartik.lakhotia@intel.com}
\affiliation{
  \institution{Intel Labs}
  \city{Santa Clara}
  \state{CA}
  \country{USA}
}

\author{Madhu Kumar}
\email{madhu.kumar@intel.com}
\affiliation{
  \institution{Intel}
  \city{Bengaluru}
  \state{Karnataka}
  \country{India}
}

\author{Sudarshan Srinivasan}
\email{sudarshan.srinivasan@intel.com}
\affiliation{
  \institution{Intel}
  \city{Bengaluru}
  \state{Karnataka}
  \country{India}
}

\author{Tushar Krishna}
\email{tushar@ece.gatech.edu}
\affiliation{
  \institution{Georgia Institute of Technology}
  \city{Atlanta}
  \state{GA}
  \country{USA}
}

\begin{abstract}

Distributed machine learning has become increasingly important due to the massive scale of large-scale generative models.
Both model parameters and data are distributed across many compute devices, which requires frequent collective communications to synchronize activations and parameter updates.
Such collective communications have become a major bottleneck.
While the performance of the collective algorithm depends on the physical network topology, the baseline collective algorithms in collective communication libraries are largely topology-agnostic.
Collective algorithm synthesizers address this inefficiency by automatically generating topology-aware collective algorithms.
However, prior works have largely overlooked that collective communication typically occurs only among a subset of devices, known as process groups.
Additionally, most existing synthesizers are limited in the range of target collective patterns they can generate.
We propose PCCL, a scalable and generic framework for synthesizing topology-aware collective algorithms.
PCCL is process group-aware and capable of generating near-optimal collective algorithms even when only a subset of devices participates in collective operations.
PCCL synthesizes arbitrary collective patterns, including 512-NPU All-to-All synthesis in 11.68~minutes.

\end{abstract}

\maketitle

\section{Introduction}\label{sec:Introduction}

Generative artificial intelligence~(AI) models have substantially increased the demand for training and serving large-scale workloads.
Large language models~(LLMs) with billions of parameters exemplify this trend~\cite{brown2020gpt3,touvron2023llama}.
The mixture-of-experts~(MoE) paradigm further facilitates massive model scaling while retaining manageable computational requirements~\cite{shazeer2017moe,cai2025moesurvey}.
For example, frontier MoE-based LLMs often comprise trillions of parameters~\cite{fedus2022switchtransformer,cerebras2024trillion}.
Consequently, the computational requirements for machine learning~(ML) model training have increased 4.7$\times$ on average per year~\cite{epoch2023aitrends}.

Given the massive scale of target workloads, training or serving large-scale ML models on a single compute device is impractical~\cite{Rajbhandari2020zero}.
This limitation has led to the widespread adoption of distributed ML approaches.
In distributed ML, multiple compute devices are employed, with each device handling a subset of the model and data~\cite{Verbraeken2020distributedmlsurvey}.
In this work, we use the term neural processing unit~(NPU) as a general abstraction encompassing various compute devices, including graphics processing units~(GPUs), tensor processing units~(TPUs), and custom ML accelerators.

The parallelization strategy defines how computations and data are distributed across the cluster.
For instance, model parallelism partitions the ML model across a subset of NPUs, while data parallelism distributes input data across multiple NPUs~\cite{Verbraeken2020distributedmlsurvey}.
As multiple NPUs process distinct shards of models and data, they must periodically synchronize their computation results, such as forward and backward pass activations and weight gradients~\cite{zhao2024adapcc}.
These synchronization steps are carried out through \emph{collective communication operations} (collectives for short)~\cite{yoo2024collapi}, as defined by the message passing interface~(MPI)~\cite{mpi2023collectives}.
Specifically, a \emph{process group}~\cite{li2020pytorchdistributed} is the set of NPUs involved in executing the collective.

\begin{footnotesize}
\begin{table}[t]
\centering
\caption{
Qualitative comparison of collective algorithm synthesizers ($\triangle$ in scalability indicates linear programming).
}
\label{table:synthesizer_comparison}

\begin{tabular}{c|c|c|c|c}
\toprule
\textbf{Synthesizer} & \textbf{Scalable} & \textbf{\begin{tabular}[c]{@{}c@{}}Generic\\ Topology\end{tabular}} & \textbf{\begin{tabular}[c]{@{}c@{}}Generic\\ Collective\end{tabular}} & \textbf{\begin{tabular}[c]{@{}c@{}}Process Group\\ Aware\end{tabular}} \\ \midrule
SCCL~\cite{cai2021sccl} &  &  & $\checkmark$ &  \\ \hline
TACCL~\cite{shah2023taccl} &  & $\checkmark$ & $\checkmark$ &  \\ \hline
Blink~\cite{wang2020blink} & $\triangle$ & $\checkmark$ &  &  \\ \hline
MultiTree~\cite{huang2021multitree} & $\triangle$ &  &  &  \\ \hline
ForestColl~\cite{zhao2025forestcoll} & $\triangle$ & $\checkmark$ &  &  \\ \hline
TACOS~\cite{won2024tacos} & $\checkmark$ & $\checkmark$ &  &  \\ \hline
TE-CCL~\cite{liu2024teccl} & $\triangle$ & $\checkmark$ & $\checkmark$ &  \\ \hline
\textbf{\begin{tabular}[c]{@{}c@{}}PCCL\\ (this work)\end{tabular}} & \textbf{$\checkmark$} & \textbf{$\checkmark$} & \textbf{$\checkmark$} & \textbf{$\checkmark$} \\ \bottomrule
\end{tabular}

\end{table}
\end{footnotesize}

\begin{footnotesize}
\begin{table}[t]
\centering
\caption{
Supported collective patterns by each synthesizer.
($\triangle$ denotes synthesis is supported but is not scalable as the solution uses an NP solution).
\ours{} further supports custom collectives, i.e., arbitrary pre/postconditions such as multicasts, point-to-point, and \calltoallv{}.
}
\label{table:collective_synthesis_comparison}

\adjustbox{max width=\columnwidth}{

\begin{tabular}{c|c|c|c|c}
\toprule
\textbf{Synthesizer} & \textbf{Reduce-Scatter} & \textbf{All-Gather} & \textbf{All-Reduce} & \textbf{All-to-All} \\ \midrule
SCCL~\cite{cai2021sccl} & $\triangle$ & $\triangle$ & $\triangle$ & $\triangle$ \\ \hline
TACCL~\cite{shah2023taccl} & $\triangle$ & $\triangle$ & $\triangle$ & $\triangle$ \\ \hline
Blink~\cite{wang2020blink} &  &  & $\checkmark$ &  \\ \hline
MultiTree~\cite{huang2021multitree} & $\checkmark$ & $\checkmark$ & $\checkmark$ &  \\ \hline
ForestColl~\cite{zhao2025forestcoll} & $\checkmark$ & $\checkmark$ & $\checkmark$ &  \\ \hline
TACOS~\cite{won2024tacos} & $\checkmark$ & $\checkmark$ & $\checkmark$ &  \\ \hline
TE-CCL~\cite{liu2024teccl} & $\triangle$ & $\triangle$ & $\triangle$ & $\checkmark$ \\ \hline
\textbf{\begin{tabular}[c]{@{}c@{}}PCCL\\ (this work)\end{tabular}} & $\checkmark$ & $\checkmark$ & $\checkmark$ & $\checkmark$ \\ \bottomrule
\end{tabular}

}

\end{table}
\end{footnotesize}

Modern data center ML clusters comprise more than 100,000 NPUs~\cite{xai2025colossus}.
As a result, collective communication among NPUs has become the most significant bottleneck in both ML training and inference~\cite{li2019bottleneck,mikami2019bottleneck,sapio2019bottleneck}.
\emph{Collective communication algorithms} define \textit{how} traffic should be routed over the physical network to execute collective communications~\cite{won2024tacos}, and the optimal collective algorithm is highly dependent on the system's underlying network topology~\cite{rashidi2022themis,gabrielyan2023topologyaware}.
This further underscores the importance of \emph{topology-aware collective algorithms}.
For this reason, many collective communication libraries~(CCLs)~\cite{nvidia2025nccl,intel2021oneccl,amd2025rccl} include customized collective algorithms tailored for their target topologies, and manually designed collective algorithms specialized for target networks have also been proposed~\cite{cho2019blueconnect,rashidi2022themis,ma2021paard,laskar2024tto}.
Unfortunately, manually designing topology-aware collective algorithms for every network topology is prohibitive, not only requiring expert knowledge but also incurring engineering and validation costs~\cite{cai2021sccl,won2024tacos}.

This has led to the development of topology-aware collective algorithm \emph{synthesizers}~\cite{cai2021sccl,shah2023taccl,wang2020blink,huang2021multitree,zhao2025forestcoll,won2024tacos,liu2024teccl}.
These synthesizers take a target topology as input and autonomously generate topology-aware collective algorithms.
We identify the following desired metrics for a collective synthesizer.

\nipara{1. Scalability}
Synthesizers must be scalable to handle the scale of ML clusters, which often comprise tens of thousands of NPUs, i.e., support topologies with many NPUs.
The pattern of collective communication calls changes frequently as the underlying network topology and target workload change~\cite{cao2025syccl}.
Even if collective synthesis is an offline process, if it takes hours for only tens of NPUs, the applicability would be limited.

\nipara{2. Supporting Generic Topologies}
Network topologies in ML clusters are often (\rom{1})~\emph{heterogeneous}, where links can vary in bandwidth and latency, and (\rom{2})~\emph{asymmetric}, where each NPU may have different numbers or shapes of connections~\cite{won2024tacos}.
Synthesizers should be able to model heterogeneous and asymmetric topologies.

\nipara{3. Supporting Generic Collectives}
Due to their use in both tensor and data parallelism, \callreduce, \creducescatter, and \callgather\ are among the most common collective patterns in distributed ML~\cite{klenk2020allreduce}.
However, MoE-based models require \calltoall\ (and sometimes \calltoallv{}) collectives for their use of expert parallelism~\cite{fedus2022switchtransformer, lei2025fast}.
These \calltoall\ communications have already emerged as a key bottleneck in MoE-based workloads~\cite{huang2023moedeployment}.

\nipara{4. Process Group Awareness}
Parallelization strategies assign parts of the workload to subsets of NPUs.
Collective communication is typically performed within these subsets, known as \emph{process groups}~\cite{li2020pytorchdistributed}.
It is uncommon for all NPUs in the cluster to participate in a single collective operation.
Therefore, it is essential for a synthesizer to recognize and exploit process group information.

\autoref{table:synthesizer_comparison} and \autoref{table:collective_synthesis_comparison} show a qualitative comparison of previously proposed collective synthesizers. Unfortunately, existing synthesizers 
fail to meet one or more of the metrics defined above.
Most of these synthesizers make trade-offs between scalability and generality of supported topologies and collective patterns.
This is an artifact of the modeling limitations of existing synthesizers: for example, integer linear programming~(ILP)-based solutions~\cite{shah2023taccl,liu2024teccl} model heterogeneous networks but are limited in scalability since ILP is a non-polynomial (NP) problem~\cite{paulus2021ilpnp}.
Spanning tree-based approaches~\cite{wang2020blink,huang2021multitree,zhao2025forestcoll} improve the scalability at the tradeoff in their heterogeneous network and switch modeling capabilities.
Moreover, none of the existing synthesizers exploit process group information; they assume that a single collective communication is executed across the entire network.
Users may employ existing synthesizers by manually defining a smaller topology of interest (i.e., a subgraph) consisting only of the NPUs in the process group.
However, this approach yields suboptimal synthesis results because the synthesizer cannot leverage available network resources outside the subgraph.
For example, ILP-based synthesizers often assume symmetric network topologies and collectives to compensate for their scalability limitations, assumptions that can be invalidated by arbitrary process groups or collectives such as multicasts or \calltoallv{}. A more comprehensive analysis is presented in \autoref{sec:RelatedWork}. 

This work aims to develop a topology-aware collective synthesizer that is both scalable and generic, while leveraging process group information.
We propose \textbf{\ours: a scalable and generic \uline{P}rocess group-aware \uline{C}ollective \uline{C}ommunica-tion \uline{L}ibrary}.
\ours\ adopts the time-expanded network~(TEN) data structure~\cite{won2024tacos}, which integrates temporal and spatial information into a unified representation.
Building on this foundation, \ours\ employs a breadth-first search~(BFS)-based pathfinding algorithm to synthesize generic collective algorithms while natively modeling heterogeneous and asymmetric networks with and without switches.  
Moreover, the BFS pathfinding algorithm naturally incorporates process group information into the synthesis process.
To summarize, this work makes the following contributions:  
\begin{itemize}
    \item We motivate the core requirements of collective algorithm synthesizers: scalability, support for generic topologies and collectives, and process group awareness.
    \item \ours\ is a collective algorithm synthesizer that satisfies all these core objectives.
    \ours\ is built on the TEN representation, which captures both temporal and spatial information of arbitrary network topologies.
    \item \ours\ introduces a BFS pathfinding algorithm that is both scalable and generic.  
    The BFS pathfinding algorithm supports arbitrary collectives, including \calltoall, and automatically incorporates process group information.  
    \item Process group-aware \ours{} synthesized an \calltoall{} algorithm 2.68$\times$ faster for \tmesh{} topology, on average.
    \item \ours{} \calltoall{} algorithm synthesis takes 11.68 minutes for an 512-NPU cluster, more than three orders of magnitude faster than a state-of-the-art synthesizer.
\end{itemize}

\section{Background}\label{sec:Background}

\insertFigure{CollectiveDefinition}{
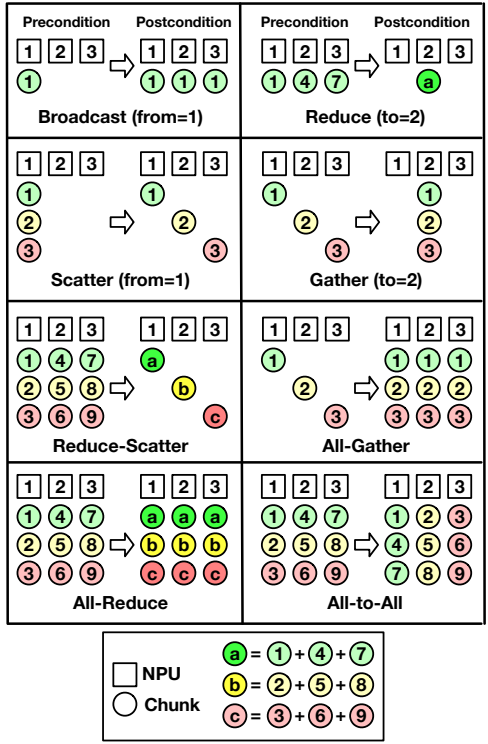
Definition of MPI collective communication patterns.
Each square denotes an NPU, whereas each circle denotes a chunk.
}{0.75}{-1em}{-1em}

\subsection{Collective Communication Pattern}\label{subsec:Background:CollectivePattern}

MPI defines several synchronization patterns among NPUs~\cite{mpi2023collectives}.
These are known as collective communication patterns and are extensively used in distributed ML clusters~\cite{yoo2024collapi,zhao2024adapcc}.
\autoref{fig:CollectiveDefinition}~illustrates the MPI collective patterns.
Each circle represents a \emph{chunk}, which is a logical unit of network transfers in collective communication.
The \emph{precondition} specifies the initial buffer contents of NPUs, while the \emph{postcondition} describes the final buffer status after the collective operation.
\calltoallv{} is a generalized version of \calltoall{} where each NPU can have different numbers of chunks in the pre/postcondition, common in MoE-based LLMs~\cite{lei2025fast}.
Some collective patterns (\creduce, \creducescatter, and \callreduce) require the reduction of chunks, typically using arithmetic addition in the context of distributed ML.

\subsection{Process Group}\label{subsec:Background:ProcessGroup}

Examples in \autoref{fig:CollectiveDefinition} are depicted by using a cluster of three NPUs, and all three NPUs are involved in the collective communication.
In practice, since parallelization strategies distribute a job across subsets of NPUs, collective communication typically runs in a more localized fashion; not all NPUs in the cluster execute a single collective communication.
Instead, a small set of NPUs in the cluster executes collective communication amongst themselves.
\emph{Process group} is the term used to denote each such set of NPUs executing a collective communication~\cite{li2020pytorchdistributed}.
As an example, \autoref{fig:ProcessGroupDefinition} shows two process groups over a six-NPU cluster.
NPUs \{1, 2, 3\} execute \creducescatter\ among themselves, forming a process group.
Another process group, composed of NPUs \{4, 5, 6\}, is running \callgather.

\subsection{Collective Communication Algorithm}\label{subsec:Background:CollectiveAlgorithm}

A \emph{collective communication algorithm} defines how each chunk should traverse (i.e., be sent and received) over the network to execute a target collective pattern~\cite{won2024tacos}.
As an example, the \aring\ \callgather\ algorithm~\cite{Thakur2005mpich} is shown in~\autoref{fig:TopologyAwareCollectivesDefinition}(a).
The \aring\ algorithm assumes NPUs are logically connected in a \tring.
Each NPU sends a chunk to its neighbor while receiving a chunk from its other neighbor.
This process is repeated for $N-1$ steps, where $N$ is the number of NPUs in the process group, until every NPU receives all chunks to satisfy their \callgather\ postcondition.
\adirect~\cite{rashidi2020astrasim}, \arhdfull~(\arhd)~\cite{Thakur2005mpich}, and \adbtfull~(\adbt)~\cite{nvidia2019dbt} are additional examples of \callreduce\ collective algorithms.

\insertFigure{ProcessGroupDefinition}{
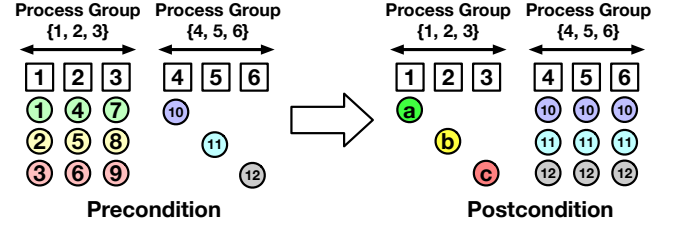
Two process groups over a six-NPU cluster.
Process group \{1, 2, 3\} is executing \creducescatter, while process group \{4, 5, 6\} is running \callgather.
(chunk $a$, $b$, and $c$ are defined in~\autoref{fig:CollectiveDefinition}).
}{1}{-2em}{-0em}

\insertFigure{TopologyAwareCollectivesDefinition}{
Examples of topology-aware and topology-unaware \callgather\ collective algorithms.
(a)~Topology-aware (unidirectional) \aring{} algorithm over a \tring{} topology.
(b)~\aring{} algorithm is not topology-aware over an example custom topology, resulting in network underutilization.
(c)~Example of topology-aware \callgather{} algorithm, showing 50\% speedup over the \aring{} algorithm.
}{1}{-2em}{-1em}

\subsection{Topology-Aware Collective Algorithm}\label{subsec:Background:TopologyAwareAlgorithm}

Note that the physical topology in~\autoref{fig:TopologyAwareCollectivesDefinition}(a) is also a unidirectional \tring.
Consequently, the \aring\ collective algorithm utilizes 100\% of the available network bandwidth and incurs no network congestion.
In this case, the \aring\ algorithm is a \emph{topology-aware collective algorithm} for the \tring\ topology---it fully leverages the physical network without network congestion, yielding optimal collective performance~\cite{won2024libra}.

In contrast,~\autoref{fig:TopologyAwareCollectivesDefinition}(b) shows the same \aring\ algorithm executed over an arbitrary physical network with eight links.
Although it performs a valid \callgather\ collective, it underutilizes the available network resources and therefore does not achieve optimal performance.
In this context, the \aring\ algorithm is \emph{not topology-aware}.
\autoref{fig:TopologyAwareCollectivesDefinition}(c) shows an example of a topology-aware \callgather\ algorithm tailored for the same custom network, achieving a 50\% speedup compared to the topology-unaware \aring\ algorithm.

\subsection{Collective Algorithm Synthesizer}\label{subsec:Background:CollectiveSynthesizer}

As discussed in~\autoref{subsec:Background:TopologyAwareAlgorithm}, executing topology-aware collective algorithms can maximize collective performance for a given target topology.
However, manually designing such algorithms is costly, requiring significant engineering and validation effort~\cite{cai2021sccl,won2024tacos}.
To address this challenge, \emph{collective algorithm synthesizers} have been proposed~\cite{cai2021sccl,shah2023taccl,wang2020blink,huang2021multitree,zhao2025forestcoll,won2024tacos,liu2024teccl}.
Rather than relying on human experts, synthesizers are automated frameworks that generate topology-aware collective algorithms.
A synthesizer takes the target network topology as input and autonomously produces optimized, topology-aware collective algorithms.

\subsection{Time-Expanded Network}

Introduced to the domain of collective communication by~\cite{won2024tacos}, a TEN is a data structure that \emph{captures both spatial and temporal information} of a network in a unified representation.
\autoref{fig:TenDefinition} illustrates an example of a TEN representation of a four-NPU cluster.
\autoref{fig:TenDefinition}(a) is the spatial layout of a four-NPU unidirectional \tring\ network.
The TEN representation of this network topology is drawn in~\autoref{fig:TenDefinition}(b).
All endpoints in a network topology comprise a column, which is then duplicated across multiple timesteps (from $t=0$ to $t=3$ in this example).
Spatial connectivities between devices are encoded as edges that span across timesteps.
TEN enables intuitive modeling of network traffic.
\autoref{fig:TenDefinition}(c) shows two such operations: chunk $a$ is sent from NPU 1 to NPU 2 at $t=0$, and chunk $b$ is sent from NPU 4 to NPU 1 at $t=2$.

\section{Motivation}\label{sec:Motivation}

In this section, we motivate the key requirements that a collective algorithm synthesizer should satisfy to be practically deployable, and define the problem statement for \ours.

\subsection{Scalability}

Modern ML clusters comprise tens to hundreds of thousands of NPUs~\cite{xai2025colossus,lee2022rsc}.  
Over such a cluster, training and inference jobs are distributed using model and data parallelism strategies.
Each model-parallel process group may include tens of NPUs~\cite{narayanan2021tpscale}, while the remaining NPUs are often organized into data-parallel process groups, resulting in a set of \emph{hundreds to even thousands of NPUs}~\cite{rae2022dpscale1,chowdhery2023palm}.
Specifically, due to the changes in model architecture, network topology, optimization techniques, and hyperparameter tuning, collective patterns issued by the workload experiences frequent changes~\cite{cao2025syccl}.
Consequently, it is critical for a synthesizer to support targets with hundreds to thousands of NPUs in a tractable time.

\subsection{Generic Topology Support}

AI supercomputers often utilize multiple networking technologies.
On-package~\cite{selig2022cs2} high-bandwidth links, NPU-to-NPU direct memory access~(DMA) links~\cite{ian2019xelink,nvidia2025nvlink,amd2020xgmi}, scale-out interconnects~\cite{adc2009ethernet,mellanox2008ib}, even photonic networks~\cite{jouppi2023tpuv4}, are all being leveraged within a single system.
Due to such diverse network technology and connectivity options, the network topologies employed in ML clusters are captured as highly asymmetric and heterogeneous~\cite{won2024tacos}.
Therefore, the synthesizer should not be limited to symmetric and homogeneous networks and should support generic topology options.

\subsection{Generic Collective Support}

\insertFigure{TenDefinition}{
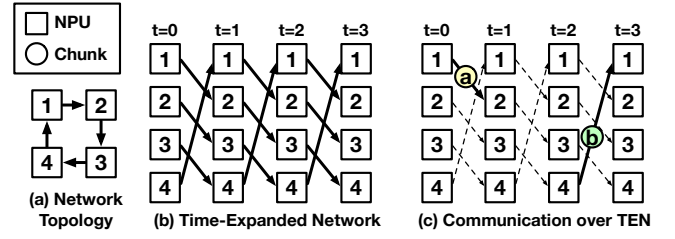
(a)~Unidirectional \tring\ topology with four NPUs. (b)~Time-expanded network~(TEN) representation of~(a), expanded up to time three. (c)~Two chunk transfer operations represented over the TEN.
}{1}{-2em}{-0.5em}

\begin{footnotesize}
\begin{table}[t]
\centering
\caption{
Collectives required by each parallelization strategy.
}
\label{table:collective_types}

\adjustbox{max width=\columnwidth}{

\begin{tabular}{c|c|c|c|c|c}
\toprule
\textbf{Parallelism} & \textbf{Reduce-Scatter} & \textbf{All-Gather} & \textbf{All-Reduce} & \textbf{All-to-All} & \textbf{Pt-to-Pt} \\ \midrule
\textbf{Data} &  &  & $\checkmark$ &  &  \\ \hline
\textbf{Tensor} & $\checkmark$ & $\checkmark$ &  &  &  \\ \hline
\textbf{Expert} &  &  &  & $\checkmark$ &  \\ \hline
\textbf{Pipeline} &  &  &  &  & $\checkmark$ \\ \bottomrule
\end{tabular}

}

\end{table}
\end{footnotesize}

Different ML model architectures and parallelization strategies require different sets of collective patterns to be executed.
Such a paradigm is summarized in~\autoref{table:collective_types}.
For example, MoE-based generative models~\cite{chowdhery2023palm,google2024gemini,rajbhandari2022deepspeedmoe} have gained their popularity since they can retain the computation requirement while massively increasing the model parameters~\cite{shazeer2017moe}, and it comes with the \calltoall\ communication cost to assign input tokens to appropriate experts~\cite{fedus2022switchtransformer}.
In fact, \calltoall\ communication takes more than 60\% of the total execution time as the cluster size increases~\cite{huang2023moedeployment}.
However, it is surprising to note that no CCLs implement specific collective algorithms for the \calltoall{} pattern~\cite{nvidia2025nccl,amd2025rccl,intel2021oneccl}.
Instead, \adirect\ (i.e., pairwise send-receive) patterns are manually implemented using the CCL's send-receive operations.

\subsection{Process Group Awareness}\label{subsec:Motivation:PG}

ML models are dispatched across the AI cluster through parallelization strategies, such as model and data parallelism.
Collective communication often runs locally within the parallelization group, known as a process group.
For example, \callreduce{} is run across data parallel groups, and \callgather{} is often executed across tensor parallel groups.
Therefore, synthesizers must reflect this and should be process group aware, rather than synthesizing collective algorithms across the entire provided cluster.

\subsection{Problem Statement}

Design a collective algorithm synthesizer that autonomously generates topology-aware collective algorithms when the network topology is provided.
Such a synthesizer should be:
\begin{itemize}
    \item Scalable to hundreds to thousands of NPUs
    \item Support generic (heterogeneous/asymmetric) topologies
    \item Covers all collective patterns, including \calltoall
    \item Synthesize collective algorithms tailored for process groups
\end{itemize}

\autoref{subsec:RelatedWork:Synthesizers} summarizes previously proposed collective algorithm synthesizers and articulates the compromises each synthesizer makes with respect to these objectives.

\section{PCCL}\label{sec:PCCL}

This section discusses how \ours\ synthesizes topology-aware collective algorithms.
Specifically, we first define the notion of a collective condition.
Then, we describe how a collective algorithm can be synthesized using a BFS pathfinding algorithm, initially targeting homogeneous networks.
Finally, we generalize the synthesis process to support switch modeling and heterogeneous networks.

\subsection{Condition}\label{subsec:PCCL:Condition}

\insertFigure{ConditionDefinition}{
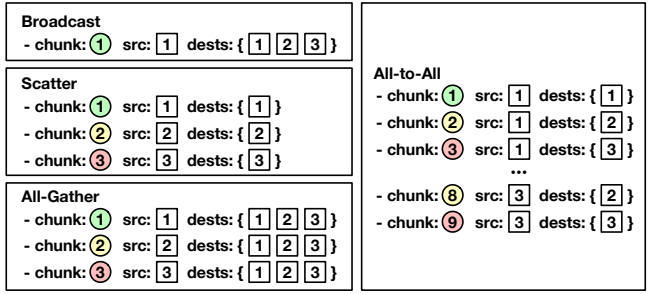
Defining collectives in~\autoref{fig:CollectiveDefinition} in a list of conditions.
Each condition defines a chunk's source NPU and destination NPUs.
}{1}{-2em}{-1em}

We first focus on collective patterns that do not require reduction operations (discussed later in~\autoref{subsec:PCCL:Reduction}), such as \callgather\ or \calltoall.
As illustrated in~\autoref{fig:CollectiveDefinition}, a chunk in these non-reduction collectives resides in only one NPU in the precondition.
In the postcondition, a chunk may reside in either a single NPU (e.g., \cscatter\ or \calltoall) or multiple NPUs (e.g., \cbroadcast\ or \callgather).

We propose defining collective communication patterns using a \emph{condition}-based representation.
Preconditions and postconditions are NPU-centric---describing which chunks each NPU holds before and after collective communication, respectively.
Meanwhile, the condition-based view is chunk-centric.
Each condition specifies a chunk’s source NPU and its set of destination NPUs.
A collective communication pattern consists of one or more collective conditions.
\autoref{fig:ConditionDefinition} shows examples of collective patterns expressed using condition-based notation.
For instance, a \cscatter\ contains three conditions, each describing a single source and destination pair for a chunk.
\callgather, on the other hand, also has three conditions, but each chunk is destined for multiple NPUs instead of just one.

\subsection{TEN Representation}\label{subsec:PCCL:TEN}

\begin{algorithm}[t]
\caption{TEN Functionality}
\label{alg:ten_functionality}

\begin{algorithmic}[1]
\Require $TEN[t][s][d]$=true: $TEN$ has a link $s \rightarrow d$ at $t$

\Function{NextDevices}{$TEN, npu, time$}
    \State \Return \{$next$: $TEN[time][npu][next]$=true\}
\EndFunction
\\

\Function{Available}{$TEN, npu, time$}
    \If{$|$\textsc{Neighbors}$(TEN, npu, time)| > 0$}
        \State \Return true
    \Else
        \State \Return false
    \EndIf
\EndFunction
\\

\Function{NextAvailableTime}{$TEN, npu, time$}
    \State $t \gets time$
    \While{not \textsc{Available}$(TEN, npu, t)$}
        \State $t \gets t + 1$
    \EndWhile
    \State \Return time
\EndFunction

\end{algorithmic}
\end{algorithm}

\begin{algorithm}[t]
\caption{BFS Pathfinding Algorithm}
\label{alg:bfs_search}

\begin{algorithmic}[1]
\Require Network $TEN$, Condition $c$
\Ensure Synthesized path $paths$ for $c$

\State $t \gets$ \textsc{NextAvailableTime}($TEN, c.src, 0$)
\State $visited \gets \{(c.src)\}$
\State $paths \gets \{c.src: []\}$

\While{$c.dests \nsubseteq visited$}
\For{$current$ in $visited$}
    \For{$next$ in \textsc{NextDevices}($TEN, current, t$)}
    \If{$next \in visited$}
        \State continue
    \EndIf

    \State Add $next$ to $visited$
    \State Add $(next:[paths[current],(t, next)])$ to $paths$\label{alg:bfs_search:grow_path}
    \EndFor
    \State $t \gets t + 1$
\EndFor
\EndWhile
\State \Return $paths$

\end{algorithmic}
\end{algorithm}

\ours\ leverages the TEN representation to synthesize collective algorithms.
In this section, we formally define the TEN structure and introduce three essential operations for processing a given TEN.
These operations are illustrated in~\autoref{alg:ten_functionality}.
TEN is a three-dimensional boolean matrix: $TEN[t][s][d]$.
A value of \texttt{true} at $TEN[t][s][d]$ indicates that there is a link from NPU $s$ to NPU $d$ at timestep $t$, meaning that NPU $s$ can initiate a chunk transfer to $d$ at that time.
Given this representation, we define the following utility procedures:

\begin{itemize}
    \item \textsc{NextDevices}($TEN$, $npu$, $time$):
    Returns the set of destination NPUs to which $npu$ can send a chunk at timestep $time$ (i.e., all $d$ such that $TEN[time][npu][d]$ is \texttt{true}).
    
    \item \textsc{Available}($TEN$, $npu$, $time$):
    Returns a boolean indicating whether $npu$ is available to send a chunk at timestep $time$ (i.e., whether there exists at least one valid destination NPU at $time$).
    
    \item\textsc{NextAvailableTime}($TEN$, $npu$, $time$):
    Returns the earliest time $t \geq time$ at which $npu$ becomes available to initiate a chunk transfer.
    For example, \textsc{NextAvailableTime}($TEN$, $npu$, 0) returns the first timestep at which $npu$ is capable of sending out a chunk.
\end{itemize}

$npus$ in these TEN functionalities are later generalized as $devices$ to accommodate network switches.
This is explained in~\autoref{subsec:switch_modeling}.

\insertFigure{ConditionBFSSearch}{
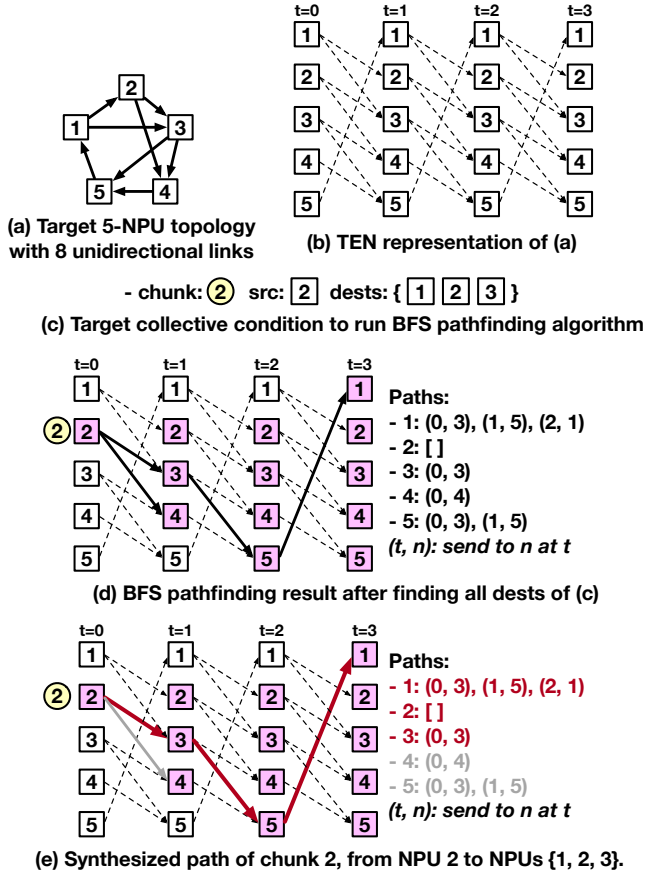
BFS search algorithm to find the path of a chunk.
(a)~Example target topology with 5 NPUs.
(b)~TEN representation of~(a), expanded up to timestep 3.
(c)~A target condition to find the route.
(d)~BFS search history to reach all destinations (NPUs \{1, 2, 3\}) of a condition.
(e)~Final chosen path of chunk 2.
}{1}{-2em}{-1em}

\subsection{BFS Pathfinding Algorithm}\label{subsec:PCCL:BFSAlgorithm}

With the TEN representation as a foundation, \ours\ determines the paths to satisfy the condition.
The objective of the BFS pathfinding algorithm is to determine the exact route for a chunk to travel from the source $c.src$ to all destinations in $c.dests$, where $c$ is a collective condition.

\ours\ represents a path as an ordered list of tuples ($t$, $n$), indicating that the chunk is sent from its current location to NPU~$n$ at timestep~$t$.
For example, consider $path=[(1, 2), (4, 3)]$ for a condition whose $c.src = 6$.
This path indicates that the chunk is first transferred to NPU~2 from NPU~6 at $t = 1$, and then from NPU~2 to NPU~3 at $t = 4$.

\ours\ pathfinding algorithm aims to construct a dictionary of such paths, $paths = \{dest: path\}$, where each entry maps the path of the chunk to reach $dest$ from $c.src$.
The pseudocode of the procedure is presented in~\autoref{alg:bfs_search}, where a condition $c$ is given and the goal is to find a path to each destination in $c.dests$.
The algorithm initializes (\rom{1})~$t$: the first available timestep for NPU $c.src$, and (\rom{2})~$visited$: the set of visited NPUs.
It then performs a BFS search over the $TEN$ to expand the $visited$ set until all $c.dests$ have been reached.
Each time a new NPU $next$ is visited, it is added to the $paths$ dictionary by appending the edge from its predecessor $npu$, building upon the path already found for $npu$, as shown in line~\ref{alg:bfs_search:grow_path} of~\autoref{alg:bfs_search}.

\autoref{fig:ConditionBFSSearch} visually illustrates this process.
\autoref{fig:ConditionBFSSearch}(a) shows an asymmetric 5-NPU network, and~\autoref{fig:ConditionBFSSearch}(b) displays the TEN representation expanded up to timestep 3.
The target condition in~\autoref{fig:ConditionBFSSearch}(c) specifies that a chunk starts at NPU 2 and must reach NPUs \{1, 2, 3\}.
\autoref{fig:ConditionBFSSearch}(d) shows the BFS traversal process: starting from source NPU 2, traversing the $TEN$ until all destinations in $c.dests$ are visited.

\begin{algorithm}[t]
\caption{Synthesizing Collective Algorithm}
\label{alg:collective_synthesis}

\begin{algorithmic}[1]
\Require Network $TEN$, Conditions Set $C$
\Ensure Synthesized Collective Algorithm $A$

\For{condition $c$ in $C$}
    \State $c.dist \gets 0$
    \For{$dest$ in $c.dests$}
        \State $c.dist \gets max(c.dist$, ShortestPath($c.src, dest$))
    \EndFor
\EndFor

\State Sort $C$ in descending order by $c.dist$
\\

\State $A \gets \{\}$
\For{$c$ in $C$}
    \State $p \gets BFS(TEN, c)$
    \State Add $p$ to $A$
    \State Remove $p$ from $TEN$
\EndFor
\State \Return $A$

\end{algorithmic}
\end{algorithm}

\insertFigureWide{FullAllGatherSearch}{
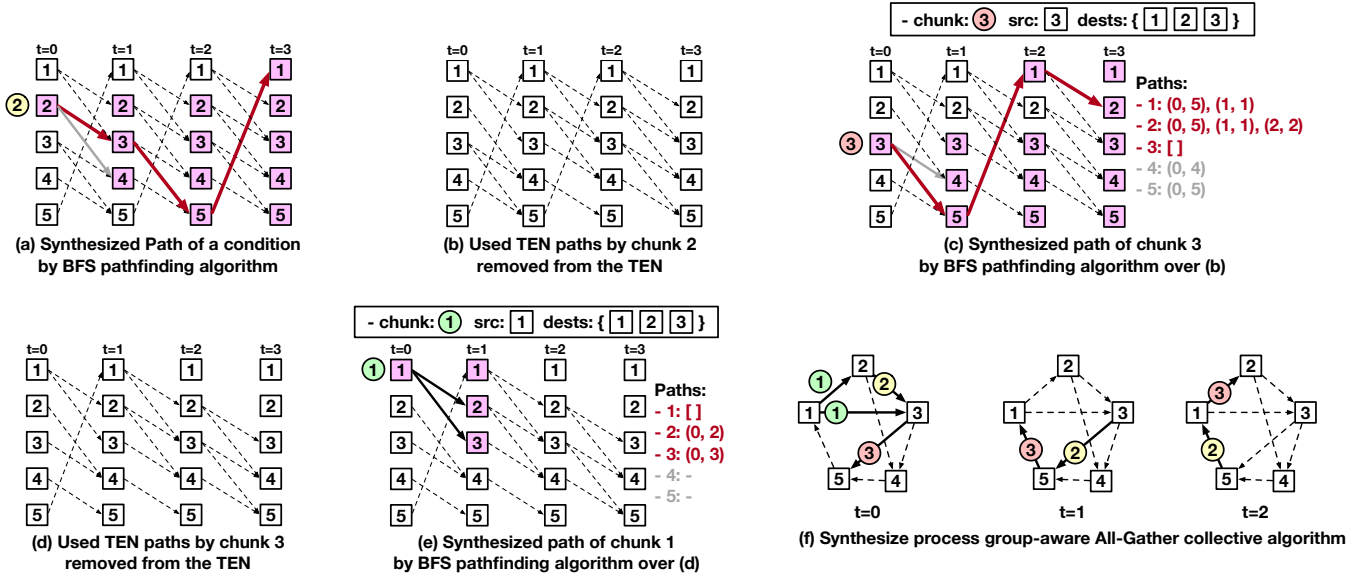
Synthesizing a \callgather\ collective algorithm for process group \{1, 2, 3\}, based on the topology shown in~\autoref{fig:ConditionBFSSearch}.
}{1}{-2em}{0em}

Glimpsed in~\autoref{fig:ConditionBFSSearch}(d) is how the BFS pathfinding algorithm may end up visiting more NPUs than the requested destinations of the condition.
For example, the BFS result in~\autoref{fig:ConditionBFSSearch}(d) visited all five NPUs, although the condition only requires visiting NPUs \{1, 2, 3\}.
Consequently, not all paths constructed during the pathfinding process are meaningful, and only the useful paths should be retained to finalize the chunk's actual path.
Simply, the process iterates over the actual destinations of the condition $c.dests$, and selects the paths associated with them.
\autoref{fig:ConditionBFSSearch}(e) shows the filtering result.
$TEN[0][2][4]$ was filtered out, since such communication is meaningless given that 4 is not in the destination set of the condition.
However, note that the chunk transfer to NPU 5 remains in the path---albeit not in the destination set.
It is because NPU 5 acts as an intermediate node to forward the chunk to NPU 1, one of the destinations.

This explains how the BFS pathfinding algorithm captures the process group information: it first tries to construct paths \emph{utilizing the entire network}, then filters out only the meaningful paths to $c.dests$.

\subsection{Synthesizing Collective Algorithm}\label{subsec:PCCL:AlgorithmSynthesis}

\autoref{fig:ConditionBFSSearch} visualized the synthesis process of a \cbroadcast\ algorithm for a chunk from NPU~2 to NPUs \{1, 2, 3\}.
However, as shown in~\autoref{fig:ConditionDefinition}, a collective pattern may consist of multiple conditions, unlike a simple \cbroadcast.
Synthesizing collective algorithms for such patterns can be achieved by repeatedly applying the BFS pathfinding algorithm.
Note that two chunks occupying the same TEN link lead to network congestion, as it indicates a conflict where multiple chunks attempt to use the same physical link at the same time.
In other words, a specific TEN link can only be occupied by a single chunk.
Therefore, to avoid link congestion in the resulting collective algorithm, any TEN links chosen during a previous BFS pathfinding step are removed from subsequent BFS searches.

The pseudocode for this process is shown in~\autoref{alg:collective_synthesis}.
Since multiple chunks must be mapped over the TEN, \ours\ must first determine which chunk should have its path synthesized first.
To assign the order, \ours\ first computes a distance $dist$ to each condition $c$, defined as the maximum shortest-path distance between $c.src$ and $c.dests$.
Then, it sorts the set of conditions $C$ in descending order of $c.dist$.
This strategy aims to maximize network resource utilization by assigning paths to chunks that must traverse the network for the longest duration first.
Chunks that traverse shorter distances can then utilize the remaining unoccupied TEN links, thereby heuristically maximizing bandwidth utilization, as motivated in~\cite{shah2023taccl}.

\autoref{fig:FullAllGatherSearch} showcases an example by extending \autoref{fig:ConditionBFSSearch} to the \callgather\ collective among a process group \{1, 2, 3\}.
\autoref{fig:FullAllGatherSearch}(a) shows the synthesized path for chunk 2, as presented in~\autoref{fig:ConditionBFSSearch}(e).
Since these links are occupied by chunk 2, they are removed from the TEN before conducting the next BFS pathfinding step, as illustrated in~\autoref{fig:FullAllGatherSearch}(b).
Subsequently, the BFS algorithm is run for chunk 3 in~\autoref{fig:FullAllGatherSearch}(c), and the resulting links are also removed from the TEN, as depicted in~\autoref{fig:FullAllGatherSearch}(d).
Note that the number of available TEN links decreases as \ours\ schedules more conditions across the network.
\autoref{fig:FullAllGatherSearch}(e) shows the path for chunk 1, obtained through an additional BFS pathfinding phase.
Finally, \autoref{fig:FullAllGatherSearch}(f) summarizes the synthesized topology-aware \callgather\ algorithm, constructed through this iterative BFS pathfinding process.

Note that \ours\ successfully synthesizes a process group-aware \callgather\ algorithm.
While \autoref{fig:FullAllGatherSearch}(f) illustrates the \callgather\ operation among NPUs \{1, 2, 3\}, the synthesized algorithm flexibly utilizes network links outside this set---for example, the links $3 \rightarrow 5$ and $5 \rightarrow 1$ are used even though NPU 5 is not part of the source or destination NPUs.
Furthermore, as \ours\ assumes no specific characteristics of a condition, it is inherently generic and can be applied to any collective pattern, including \calltoall.

\insertFigure{ReductionOperations}{
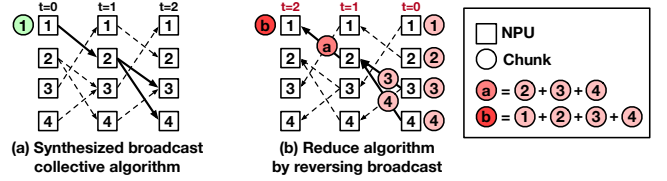
(a)~Example \cbroadcast\ collective algorithm for a 4-NPU cluster.
(b)~Synthesized \creduce\ operation by reversing the directions of~(a) and applying reduction operations.
}{1}{-2em}{-1em}

\subsection{Reduction Operations}\label{subsec:PCCL:Reduction}

Collective patterns involving reductions can be supported by synthesizing their corresponding non-reduction collective algorithms, following the paradigm introduced in~\cite{cai2021sccl,shah2023taccl,won2024tacos}.
\autoref{fig:ReductionOperations} illustrates this approach.
To synthesize a \creduce\ algorithm, \ours\ first generates the corresponding \cbroadcast\ algorithm, as shown in~\autoref{fig:ReductionOperations}(a).
By reversing the direction of all transfers and applying reduction operations, the \creduce\ algorithm can be automatically constructed, as depicted in~\autoref{fig:ReductionOperations}(b).
Similarly, \creducescatter\ can be synthesized by reversing \callgather, and \callreduce\ is realized by composing \creducescatter\ followed by \callgather~\cite{shah2023taccl}.

\subsection{Heterogeneous Networks}

\ours\ leverages the $\alpha$-$\beta$ network model~\cite{hockney1994alphabeta} to support heterogeneous networks, as suggested in~\cite{shah2023taccl,won2024tacos}. 
The $\alpha$-$\beta$ model estimates the transfer time of a link as $\alpha + (m \times \beta)$, where $\alpha$ represents the link latency, $\beta$ is the reciprocal of the link bandwidth, and $m$ is the message size (i.e., chunk size for the synthesizers).
\autoref{fig:AlphaBetaModel}(a) depicts a heterogeneous network topology with two distinct links.
For such a network, the $\alpha$-$\beta$ model is used to capture the transfer time of each link as a single value, as illustrated in~\autoref{fig:AlphaBetaModel}(b).
\autoref{fig:AlphaBetaModel}(c) shows the TEN representation corresponding to~\autoref{fig:AlphaBetaModel}(b), where the timesteps in $\mu$s are from the $\alpha$-$\beta$ model.

\insertFigure{AlphaBetaModel}{
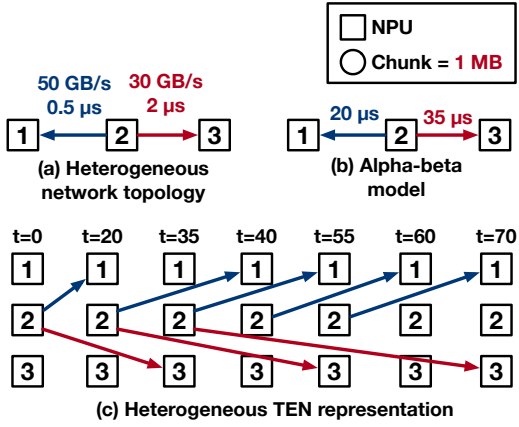
(a)~A heterogeneous network with two links of different bandwidths and latencies.
(b)~Application of the $\alpha$-$\beta$ model with a chunk size of 1\,MiB.
(c)~TEN representation of~(b).
Note that the timesteps reflect the timing information from the $\alpha$-$\beta$ model.
}{0.8}{-1em}{-1em}

For heterogeneous TEN representations, \ours{}'s BFS pathfinding is applicable with only marginal modifications. 
Such modifications mostly arise from the timing considerations when a chunk arrives at the next NPU, and which TEN links to disable to avoid network congestion.

\nipara{Chunk Arrival Time}
Since each link can have a different transfer time, \ours\ must track the exact arrival time of each chunk at every device.
As a result, the $visited$ set is extended to include the $time$ information representing the timestep at which the chunk reaches each device.
During the BFS process, whenever a $current$ device is processed, it must be ensured that the chunk has actually arrived at this source.

\nipara{Removing TEN Links}
Additional care must also be taken when removing TEN links during subsequent BFS pathfinding passes.
When a TEN link is used, not only that specific link, but also all other links overlapping with its timestep must be removed to avoid network congestion.
This scenario is illustrated in~\autoref{fig:HeterogeneousLinkRemoval}.
If a chunk is sent at $t=1$ from NPU~1 to 2, $TEN[0][1][2]$ and $TEN[2][1][2]$ are also disabled to avoid network congestion.

\insertFigure{HeterogeneousLinkRemoval}{
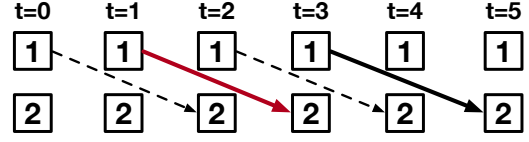
Removing TEN links in a heterogeneous network.
If a TEN link from $t=1$ to $t=3$ is taken, other TEN links overlapping with this timestep (e.g., $TEN[0][1][2]$ and $TEN[2][1][2]$) must be disabled to prevent network congestion.
}{0.8}{0em}{-1em}

\subsection{Modeling Switches}\label{subsec:switch_modeling}

Switch modeling remains an open question in collective synthesizers today.
Most past works unroll a switch into direct-connect links\cite{shah2023taccl,won2024tacos,zhao2025forestcoll}.
Unfortunately, this limits modeling switch-specific considerations (e.g., finite buffers, or unicast versus multicast support).
In \ours{}, while we inherently support unrolling, we also add explicit support to model switches via two classes of TEN nodes: NPUs and switches. We track the number of chunks ``buffered" at each switch node, and the BFS pathfinding algorithm skips visiting a TEN node at a given timestep if the switch node exceeds the provided buffer size. Furthermore, if the switch node does not support multicast, the BFS algorithm visits only one next neighbor when the node type is a switch. Even under this restriction, as the BFS algorithm visits other nodes in subsequent timesteps (i.e., rather than visiting all next neighbors at once, it visits next nodes one by one), the algorithm can still successfully synthesize a collective algorithm.

\subsection{Translating Synthesis Results}

It is important to note that \ours{} does not specifically target GPU-centric networking, but rather contributes to the algorithmic foundations for synthesizing arbitrary collective algorithms at scale.
Nevertheless, this section illustrates that \ours{} synthesis results can be translated into other representations to execute on a specific target system.
For GPU-based systems, as an example, we propose to use \fmsccl{}~\cite{cowan2023mscclang} and \fmscclpp~\cite{shah2025mscclpp} representations.
Both frameworks enable the usage of scratch buffers and multiple threadblocks per GPU for concurrent communications.
For further optimization, \fmscclpp{} even allows a single threadblock to concurrently put messages to multiple peers.
Meanwhile, \fmsccl{} not only provides \texttt{send} and \texttt{receive} operations but also compound operations such as \texttt{receive-copy-send}, all suitable to represent \ours{} chunk operations.
Therefore, by leveraging \fmsccl{} and \fmscclpp{}, \ours{} synthesis results can be readily represented and executed on GPU systems without any modifications to the synthesizer itself.

\section{Methodology}\label{sec:Methodology}

\subsection{Experimental Infrastructure}

We use ASTRA-sim, a distributed ML systems simulator for the evaluation of this work~\cite{won2023astrasim2,rashidi2020astrasim}, similar to the setup of~\cite{won2024tacos}.
ASTRA-sim simulation has been validated over a real system comparison using a 128 NVIDIA H100 cluster with the accuracy of 97\%~\cite{astrasim2025validation}.
We also validated the correctness of the \ours{}-synthesized algorithm by executing the MSCCL-IR over 16 and 32-GPU clusters via the MSCCL executor~\cite{cowan2023mscclang}.

\subsection{Baseline Collectives}

We mostly evaluate \ours{} by targeting the \calltoall{} collective pattern, as (\rom{1})~most synthesizers fail to synthesize such an algorithm, especially at scale, and (\rom{2}) no collective algorithm exists in CCLs.
\ours{} mechanism is still applicable to generic collective patterns and to showcase that, we also have \calltoallv{} and \callgather{} results as well.
We use \adirect{}, a pairwise point-to-point send-receive algorithm, as the baseline collective algorithm to compare \calltoall{} performance, as such a mechanism is what CCLs use today.

\section{Results}\label{sec:Results}

In this section, we evaluate the performance and efficacy of the \ours{} synthesizer.
Specifically, we emphasize checking the four practical objectives for collective synthesizers discussed in~\autoref{sec:Motivation}: scalability, generic topology support, generic collective support, and process group awareness.

\insertFigure{npu_size_scalability}{
\calltoall{} synthesis time of \ours{} vs. \steccl{} time for small \tmesh{}.
For a small 6$\times$6 (36 NPU) topology, \ours{} is already 4,404$\times$ faster over \steccl{}.
Further topology scalability analysis of \ours{} is also shown using \thypercube{}.
}{0.9}{-1em}{-1em}

\insertFigure{coll_size_sensitivity}{
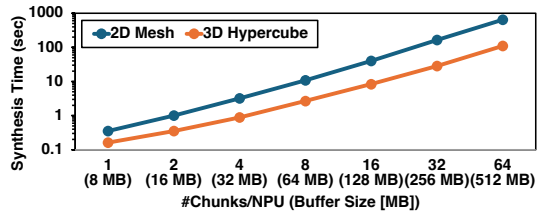
\ours{} synthesis time of \calltoall{} algorithm (8--512\,MiB) for 64-NPU 2D Mesh and 3D Hypercube topologies, by increasing the number of 128\,KiB chunks per each NPU.
}{0.85}{-1em}{-1em}

\subsection{Scalability}\label{subsec:scalability}

Firstly, we evaluate the scalability of the \ours{} synthesizer.
For the scalability analysis, we evaluated \calltoall{} collective pattern as it has the largest search space and thereby most synthesizers struggle to synthesize.
Therefore, we mainly target \steccl{}~\cite{liu2024teccl}, the state-of-the-art in \calltoall{} synthesis, for main scalability comparison.

\nipara{Topology Size}
We measured the \calltoall{} synthesis time of \ours{} by increasing the size of \tmesh{} and \thypercube{} target topologies.
\autoref{fig:npu_size_scalability} summarizes the observed synthesis time.
Notably, only at 36-NPU cluster scale, \ours{} was already more than 3 orders of magnitude faster than the state-of-the-art \steccl{}.
Further scaling the target topology shows that \ours{} can synthesize an \calltoall{} algorithm for a 512-NPU cluster in just 11.68 minutes, and 1,000-NPU cluster in 2.01~hours.
The complexity to synthesize \calltoall{} algorithm was $O(n^3)$.
Specifically, we measured TE-CCL taking 3~minutes for a 36-NPU (6$\times$6 Mesh) target and more than 30 minutes for 49 NPUs.
Although TE-CCL was able to synthesize a 256-GPU target in 25 minutes~\cite{liu2024teccl}, Cao et al.~\cite{cao2025syccl} report that TE-CCL takes 4.4–49.5~minutes and 3.8~minutes–8.7~hours for 16- and 32-GPU systems, respectively.
This indicates that TE-CCL synthesis time heavily depends on the synthesizer setup and hyperparameters, including the search policy, number of chunks per collective, and target network topologies.
Given this large variability, we expect \ours{} to demonstrate consistently better scalability than optimizer-based synthesizers.

\nipara{Collective Size}
We also measured the scalability of the target collective pattern.
Specifically, we synthesized \calltoall{} algorithms for an 8$\times$8 Mesh.
The algorithm buffer size spans 8--512\,MiB by fixing each chunk size to 128\,KiB and increasing the number of chunks per NPU from 1 to 64.
The synthesis time is summarized in~\autoref{fig:coll_size_sensitivity}.
Notably, \ours{} synthesized a 512\,MiB \calltoall{} algorithm for a \thypercube{} topology in 1.83 minutes.
As this experiment is done by setting each chunk size to 128~KiB, the synthesis time can further be decreased by increasing the chunk size and reducing the number of chunks per NPU.

\insertFigure{switch_2d_a2a}{
\calltoall{} bandwidth of \ours{} vs. CCLs and collective speedup over heterogeneous 2D Switch topology.
Each node size is 8~NPU, and the network size spans 16--256~NPUs by increasing the number of nodes in the cluster.
}{0.85}{-1.5em}{-1em}

\insertFigure{mesh_a2a}{
Normalized \calltoall{} bandwidth when the entire \tmesh{} cluster is executing a \calltoall{} collective.
}{0.95}{-1em}{-1em}

\subsection{Supporting Generic Topology}

In~\autoref{subsec:scalability}, we already evaluated two homogeneous, asymmetric topologies: \tmesh{} and \thypercube{}.
In this section, we show the applicability of \ours{} towards more classes of topologies: notably, 2D Switch topology, which is heterogeneous.
\autoref{fig:switch_2d_a2a} showcases the same experiment by using a heterogeneous 2D switch topology: each node size is 8, and the cluster size spans 16--256 NPUs by increasing the number of nodes in the cluster.
For this setup, \ours{} showed consistent speedup over the baseline CCLs, 1.33$\times$ on average.

\subsection{Supporting Generic Collective}

In this section, we explain the applicability of \ours{} for generic collective patterns. First, we evaluate the performance of \calltoall{} pattern, with the largest search space, in more detail. \autoref{fig:mesh_a2a} compares the normalized \calltoall{} bandwidth of \ours{}, CCLs, and state-of-the-art \steccl{} synthesizer.
For small-sized Meshes, \ours{} and \steccl{} showed comparable performance.
However, \steccl{} synthesis breaks after 5$\times$5 Mesh since it is set to search for the very first satisfiable \calltoall{} algorithm due to scalability considerations, which, even after this setup, is 3 orders of magnitude slower than \ours{} (explained in~\autoref{subsec:scalability}).
After 7$\times$7 Mesh, \steccl{} synthesis takes more than 30 minutes (explained in~\autoref{subsec:scalability}), and \ours{} continuously shows better performance than the baseline CCL algorithm.

\subsection{Synthesis with Process Group}

\insertFigure{ProcessGroupSynthesisResult}{
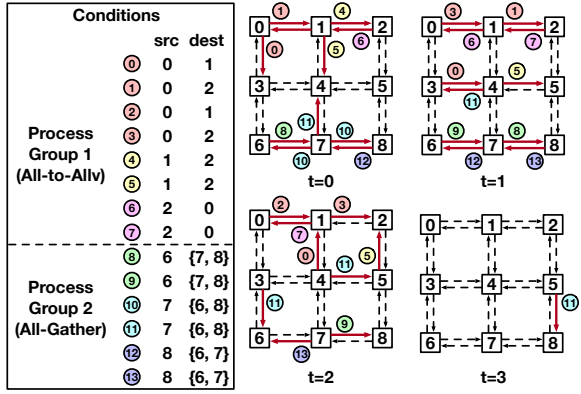
\ours{}-synthesized collective algorithm of two process groups, running \calltoallv{}~(NPUs 0--2) and \callgather{}~(NPUs 6--8).
NPUs 3--5 network resources are actively used, even though they are not in any of the process group.
}{0.9}{-1em}{-0.5em}

\insertFigure{process_group_bandwidth}{
Normalized \calltoall{} bandwidth of \ours{}-synthesized algorithm over the baseline \adirect{}, as the \tmesh{} size and the number of process groups gradually increase.
}{0.85}{-1.5em}{-1em}

\insertFigure{process_group_heatmap}{
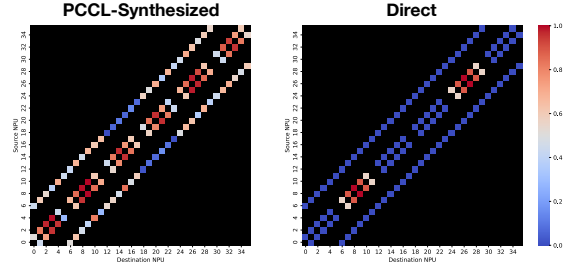
Normalized link utilization heat map of \ours{}-synthesized vs. \adirect{} collective algorithms, when two process groups are executing \calltoall{} amongst them.
Unlike \ours{}-synthesized \calltoall{} algorithm, \adirect{} fails to leverage the entire network outside the process group, resulting in 2.8$\times$ speedup.
}{0.95}{-1.5em}{-0.5em}

\insertFigure{utilizationGraph}{
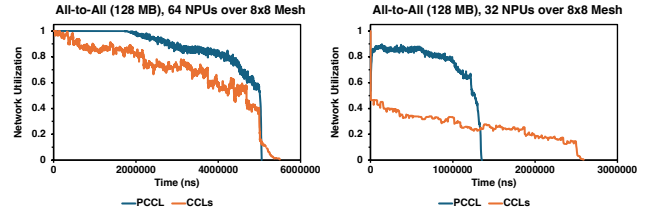
Network bandwidth utilization over time, when running 128\,MiB \calltoall{} collective over an 8$\times$8 \tmesh{}, with processing group of size 64 and 32, respectively.
}{1}{-2em}{-1em}

We then showcase the capabilities of \ours{} with a simple synthesis example.
Over a 3$\times$ Mesh, we overlaid two process groups of size three each: group 1 running \calltoallv{} (NPUs 0–2, with NPU 0 transmitting twice as much data as NPUs 1–2), and group 2 executing \callgather{} (NPUs 6–8), with two chunks per collective.
The synthesis result is depicted in~\autoref{fig:ProcessGroupSynthesisResult}.
\ours{} generated a congestion-free collective algorithm over an asymmetric \tmesh{} topology.
\ours{} also supports arbitrary collectives, provided that pre-/postconditions are specified, as illustrated in this example with \calltoallv{}.
Finally, note the process group awareness of \ours{}: (\rom{1})~multiple process groups running independent collectives are naturally supported, and (\rom{2})~NPUs 3–5 and their associated links, even though not part of any process group, are actively leveraged by the \ours{}-generated collective algorithm to maximize performance and resource utilization.

We also evaluate the benefits of process group-aware \ours{} through \calltoall{} collective.
A large ML cluster with many NPUs often follows this scheme, where multiple groups of collectives are concurrently run, rather than the entire cluster executing a single collective communication.
We measured the \calltoall{} bandwidth of pairwise \adirect{} algorithm (namely CCLs) as well as \ours{}-generated algorithms.
In doing so, we increased both the size of the target \tmesh{} topology as well as the concurrent numbers of process groups (we set the process group size equal to the Mesh width in this experiment).
The normalized algorithmic bandwidth is summarized in~\autoref{fig:process_group_bandwidth}.
Notably, \ours{}-synthesized algorithm showed 2.33--3.03$\times$ speedup over the baseline \adirect{} (2.68$\times$ average).

This can be explained by the link utilization heat map captured in~\autoref{fig:process_group_heatmap}.
When there are two process groups running \calltoall{} within themselves, \ours{} still leverages the entire network resources to maximize the performance of both \calltoall{} executions.
However, the traffic pattern generated by the \adirect{} algorithm only utilizes localized network resources, resulting in huge network underutilization.
The same is applicable to all other previous collective algorithm synthesizers, as none of the synthesizers consider process group and only generate localized collective algorithms.
This is further explained in~\autoref{fig:utilizationGraph}.
Even when running \calltoall{} over an entire topology (64 NPUs over an 8$\times$8 \tmesh{}), \ours{} still outperformed in terms of network resource utilization over the baseline \adirect{} algorithm, showcasing shorter collective time.
However, even when the process group is smaller than the topology, unlike the baseline which underutilizes the network, \ours{} still maximizes resource utilization, finishing the collective 1.88$\times$ faster.

\insertFigure{pg_scalability}{
Normalized \calltoall{} bandwidth over CCLs, when the number of 128\,MiB \calltoall{} process groups of size 8 increases over an 8$\times$8 Mesh topology.
}{0.85}{-1em}{-1.5em}

The same trend is also exemplified by a sensitivity analysis shown in~\autoref{fig:pg_scalability}.
Here, we fixed the target topology to 8$\times$8 Mesh, and varied the number of concurrent 128\,MiB \calltoall{} process groups (each of size 8).
Since \ours{} can leverage a lot of free network resources, when there was only one process group executing the collective, \ours{} showed 3.05$\times$ speedup.
As more network resources start to conflict across process groups, the benefit of \ours{}-synthesized algorithm decreases as the concurrent process group increases.

\section{Related Work}\label{sec:RelatedWork}

\subsection{Collective Algorithm Synthesizers}\label{subsec:RelatedWork:Synthesizers}

Existing synthesizers fall into three categories based on their synthesis strategy.

\nipara{Tree-based}
A spanning tree provides the routing structure for reducing and broadcasting data~\cite{Thakur2005mpich}, so tree-based synthesizers generate such trees for the target topology.

\begin{itemize}
\item \sblink~\cite{wang2020blink} constructs multiple disjoint spanning trees, enabling parallel chunk processing.
It uses LP to maximize the number of disjoint trees.
LP dependence limits scalability---despite polynomial theoretical complexity, practical solvers scale superlinearly~\cite{xiong2025lpcomplexity}.
\sblink{} only applies to direct-connect topologies and \callreduce{}.

\item \smultitree~\cite{huang2021multitree} improves generality by generating a spanning tree rooted at each NPU, supporting \callgather{} and \creducescatter{}.
Its greedy construction is highly scalable but restricted to homogeneous topologies.

\item \sforestcoll~\cite{zhao2025forestcoll} similarly generates per-NPU spanning trees and extends support to switch-based topologies via an LP transformation that replaces switches with multiple direct links.
This enables limited switch modeling but inherits LP scalability constraints and applies only to symmetric networks.
\end{itemize}

\nipara{Optimizer-based}
Tree-based synthesis provides scalability but limited generality.
Optimizer-based methods assign chunk paths using global optimization, supporting arbitrary collectives including \calltoall, at the cost of NP-complete formulations~\cite{paulus2021ilpnp}.

\begin{itemize}
\item \ssccl~\cite{cai2021sccl} uses satisfiability modulo theories~(SMT) to express collective constraints, supporting all patterns but with poor scalability due to NP-hardness~\cite{rakadjiev2015smtnp}.
It assumes homogeneous, symmetric, switch-free topologies.

\item \staccl~\cite{shah2023taccl} replaces SMT with ILP and supports heterogeneous networks.
Switches are again substituted with multiple direct links.
ILP’s NP-hardness restricts scalability; e.g., \staccl{} fails on a 16-NPU \calltoall{} within 30 minutes.

\item \steccl~\cite{liu2024teccl} uses a multi-commodity flow model that natively represents switch devices.
It reduces \calltoall{} synthesis to LP, yielding moderate scalability for this pattern, but retains ILP for other collectives.
For example, \steccl{} synthesizes a 128-NPU \calltoall{} in 43 minutes but requires over 350\,GiB of memory.

\end{itemize}

\nipara{Greedy-based}
Greedy synthesizers maximize scalability by heuristically selecting each chunk’s path.

\begin{itemize}
\item \stacos~\cite{won2024tacos} uses the TEN representation and greedily matches chunks to links.
It is highly scalable but supports only \callgather, \creducescatter, and \callreduce.
Switches are modeled using the same indirect substitution approach as \sforestcoll{} and \staccl.
\end{itemize}

Overall, prior synthesizers ignore process group structure and assume whole-cluster collectives.
They also face trade-offs between scalability and generality.

\subsection{Customized Collective Algorithms}

Several works manually design topology-specific collective algorithms.
\cablueconnect~\cite{cho2019blueconnect} and \cathemis~\cite{rashidi2022themis} target symmetric multi-dimensional networks.
\catto~\cite{laskar2024tto} and \capaard~\cite{ma2021paard} specialize \callreduce{} for \tmesh{} and \tdragonfly~\cite{kim2008dragonfly}, respectively.
\fmscclang~\cite{cowan2023mscclang} and \fmscclpp{}~\cite{shah2025mscclpp} provide DSLs for manually specifying such algorithms.
These approaches require substantial engineering effort and lack portability across diverse or evolving topologies.
In contrast, synthesizers like \ours{} automatically generate optimized collectives for a given network.

\section{Conclusion}\label{sec:Conclusion}

Designing and executing topology-aware collective algorithms is pivotal to optimizing collective communication, which is a major bottleneck in distributed ML.
This paper proposes \ours, a scalable and generic collective algorithm synthesizer.
\ours\ leverages a BFS pathfinding algorithm over a TEN, resulting in a scalable, generic, and process group-aware synthesizer.

\bibliographystyle{ACM-Reference-Format}
\bibliography{refs}

\end{document}